\documentclass[12pt]{article}
\usepackage{axodraw}
\usepackage{epsfig}
\usepackage{latexsym}
\usepackage{amssymb}
\usepackage{amsfonts}
\newcommand{\E}{\mbox{e}}
\newcommand{\be}{\begin{equation}}
\newcommand{\ee}{\end{equation}}
\newcommand{\bea}{\begin{eqnarray}}
\newcommand{\eea}{\end{eqnarray}}
\newcommand{\al}{\alpha}

\newcommand{\gm}{\gamma}
\newcommand{\Gm}{\Gamma}
\newcommand{\dl}{\delta}

\newcommand{\ep}{\varepsilon}
\newcommand{\sg}{\sigma}

\newcommand{\dd}{\mbox{d}}

\newcommand{\nn}{\nonumber}

\newcommand{\I}{i}
\newcommand{\insl}{\not\!\, \in}

\newcommand{\ual}{\underline{\alpha}}

\begin{document}
\parindent=1.5pc

\begin{titlepage}

\bigskip
\begin{center}
{{\large\bf
Hepp and Speer Sectors within Modern Strategies of Sector Decomposition
} \\
\vglue 5pt
\vglue 1.0cm
{\large  A.V. Smirnov}\footnote{E-mail: asmirnov80@gmail.com}\\
\baselineskip=14pt
\vspace{2mm}
{\normalsize Scientific Research
Computing Center of Moscow State University}\\
\baselineskip=14pt
\vspace{2mm}
and\\
\baselineskip=14pt
\vspace{2mm}
{\large   V.A. Smirnov}\footnote{E-mail: smirnov@theory.sinp.msu.ru}\\
\baselineskip=14pt
\vspace{2mm}
{\normalsize
Nuclear Physics Institute of Moscow State University\\
}
\baselineskip=14pt
\vspace{2mm}
\vglue 0.8cm
{Abstract}}
\end{center}
\vglue 0.3cm
{\rightskip=3pc
 \leftskip=3pc
\noindent
Hepp and Speer sectors were successfully used in the sixties and seventies
for proving mathematical theorems on analytically or/and
dimensionally regularized and renormalized Feynman integrals at Euclidean
external momenta. We describe them within recently developed strategies of
introducing iterative sector decompositions. We show that Speer sectors
are reproduced within one of the existing strategies.
\vglue 0.8cm}
\end{titlepage}

\section{Introduction}

The so-called alpha representation [1--11]
was initially used to introduce dimensional regularization \cite{dimreg1,dimreg2}
and to prove
various mathematical results on Feynman integrals \cite{Hepp,Speer1,Speer2,Bergere,BM,Po,books1,books1a}.
The standard way to analyze convergence of Feynman integrals is to
decompose the initial integration domain of alpha parameters into appropriate
subdomains (sectors) and introduce, in each sector, new variables in such a way
that the integrand factorizes, i.e. becomes equal to a monomial in new
variables times a non-singular function. This procedure turned out to be
successful for Euclidean external momenta, i.e with $\left(\sum q_i
\right)^2<0$ for any partial sum, when the sectors of Hepp \cite{Hepp} and Speer
\cite{Speer2} were introduced.

However, in practice, one often deals with Feynman integrals on a
mass shell or at a threshold. In this case, Hepp or Speer sectors generally do
not provide a factorization of the integrand so that the analysis of
convergence fails within this technique. Therefore, general theorems on such
`physical' Feynman integrals could not be proved up to now.

Recently Binoth and Heinrich introduced sector decompositions of a new kind
\cite{BH}. (See \cite{Heinrich} for a review.) They provided a powerful
method of evaluating Feynman integrals numerically in situations with severe UV, IR and
collinear divergences.
In contrast to Hepp and Speer sectors, the sectors of \cite{BH} are introduced
iteratively, according to so-called \textit{sector decomposition strategies}.
The corresponding algorithm was implemented on a computer.
Although this algorithm was successfully applied to
numerically evaluate complicated Feynman integrals and to check
existing analytical results (see, e.g. \cite{3box,3loop,4loop}) it was unclear where
this iterative procedure stops at some point, i.e. results in the factorization of the
integrands so that one can apply it for numerical evaluation.
Indeed, in some examples, closed loops appear within this algorithm.

The first algorithm guaranteed to terminate was developed by Bogner and
Weinzierl \cite{BognerWeinzierl}. More precisely, certain strategies within
this algorithm guarantee that closed loops do not appear.
The algorithm works at least if squares $\left(\sum q_i\right)^2$
of partial sums of the external momenta are either negative or zero.\footnote{
Let us stress that Hepp and Speer sectors generally do not provide
a resolution of the singularities in the parameter of dimensional
regularization if some sum $\sum q_i$ is light-like.}
(We will formulate a more general condition in the next section.)
The corresponding computer code is public.
The second public algorithm \cite{FIESTA}
called {\tt FIESTA} provided one more strategy
(Strategy~S) for sector decompositions which also leads to a factorization of
the integrand for general Feynman integrals. It was successfully applied in
\cite{FIESTA-appl}.

The purpose of this paper is to describe Hepp and Speer sectors in an iterative
way, within the modern technique of sector decompositions. Using examples of
three- and four-loop diagrams, we will also see
that Speer sectors are reproduced within Strategy~S. In the next section, we
describe the alpha representation and Hepp and Speer sectors in a simplified
form. Then, in Section~3, we establish their connection with iterative sector
decomposition. Finally, we discuss some perspectives.

\section{Hepp and Speer sectors}

For a Feynman integral with standard propagators
(of the $1/(m^2-k^2-i0)^{a_l}$ form)
corresponding to a connected graph $\Gm$, the alpha representation has the following form:
\bea
F_{\Gm}(q_1,\ldots,q_n;d;a_1\ldots,a_L)
& =&
\frac{\I^{a+h(1-d/2)}  \pi^{h d/2}}{\prod_l \Gm(a_l)}
\nn \\ && \hspace*{-30mm}
\times \int_0^\infty\ldots\int_0^\infty
\prod_l \al_l^{a_l-1}
{\cal U}^{ -d/2}   \E^{\I {\cal V}/{\cal U}-\I\sum m_l^2 \al_l}
\dd\al_1 \ldots \dd\al_L \; ,
\label{alpha}
\eea
where  $L$ and $h$ is, respectively, the number of lines (edges) and loops
(independent circuits) of the graph, $n+1$ is the number of
external vertices,
$a=\sum a_l$, and
\bea
{\cal U}& = & \sum_{T\in T^1} \prod_{l\insl  T} \al_l
 \;,
\label{Dform}  \\
{\cal V}&=& \sum_{T\in T^2} 
\prod_{l\insl  T} \al_l
\left( q^T\right)^2
 \; .
\label{Aform}
\eea
In (\ref{Dform}), the sum runs over
trees
of the given graph,
and, in (\ref{Aform}),
over {\em 2-trees}, i.e. maximal subgraphs that do not involve loops
and consist of two connectivity components; $q^T$ is
the  sum of the external momenta that flow
into one of the connectivity components of the 2-tree $T$. (It does not
matter which component is taken because of the conservation law for the
external momenta.) The products of the alpha parameters involved are taken
over the lines that do not belong to the given (2-)tree $T$.
The functions $\cal U$ and $\cal V$ are homogeneous functions
of the alpha parameters with the homogeneity degrees $h$ and
$h+1$, respectively.

An overall integration can be performed to obtain another
well-known parametric representation\footnote{
So, the code of \cite{BognerWeinzierl} works at least if each monomial of the
$\al$-variables in the function $-{\cal V} +{\cal U}\sum m^2_l \al_l$ enters with
a negative coefficient. This condition is a little
bit relaxed within the code of \cite{FIESTA} where combinations of the type
$(\al_i-\al_j)^2$ are also admissible.}:
\bea
F_{\Gm}(q_1,\ldots,q_n;d;a_1\ldots,a_L)
& =&
 \frac{\left(\I\pi^{d/2} \right)^h
\Gm(a-h d/2)}{\prod_l\Gm(a_l)}
\nn \\ && \hspace*{-56mm}
\times
\int_0^\infty \ldots\int_0^\infty  \prod_l \al_l^{a_l-1}\,
\dl\left( \sum \al_l-1\right)
\frac{
{\cal U}^{a-(h+1) d/2}}
{  \left(-{\cal V} +{\cal U}\sum m^2_l \al_l\right)^{a-h d/2}}
\dd\al_1 \ldots \dd\al_L \;.
\label{alpha-d-mod}
\eea
According to the well-known folklore Cheng--Wu theorem one can choose any sum
of the alpha variables in the argument of the delta function.

We imply that the graph $\Gm$ is a {\em connected} graph, i.e.
any two vertices of $\Gm$ can be connected by a path in $\Gm$. However,
we are going to consider various subgraphs of the graph and they
can be disconnected, i.e. consist of several connectivity components.
A subgraph $\gm$ of $\Gm$ is determined by a
subset of lines ${\cal L}(\gm)$ and includes all the vertices
incident to these lines. (Sometimes isolated vertices are added to a subgraph.
For example, {\tt Mathematica} produces isolated vertices
as bi-connected components.) The number of loops  of a
subgraph equals to
\[
h(\gm)=L(\gm)-V(\gm)+c(\gm)\;,
\]
where $V(\gm)$ and $c(\gm)$ are, respectively the numbers of the
vertices and connectivity components.


\textit{An articulation vertex} of a graph $\Gm$ is a vertex whose deletion 
disconnects $\Gm$. Any graph with no articulation vertices is said to
be \textit{biconnected} (or, \textit{one-vertex-irreducible} (1VI)).
Otherwise, it is called \textit{one-vertex-reducible} (1VR).
In other words, in a 1VR graph, one can distinguish two
subsets of its lines and a vertex (an articulation vertex)
such that any path between vertices from
these two subsets goes through this vertex.
From now on let us suppose that we are dealing with a 1VI graph.
It is natural to treat a single line 
as a 1VI graph since we cannot decompose it into two parts.

\begin{figure}[htb]
\begin{center}
\includegraphics[]{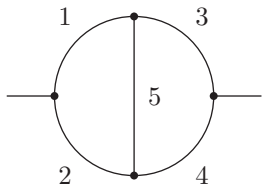}
\caption[]{A two-loop propagator diagram}
\label{fi14}
\end{center}
\end{figure}

Any subgraph can be represented as the union of its 1VI
components, i.e. maximal 1VI subgraphs.
Consider, for example,  the two-loop self-energy diagram of
Fig.~\ref{fi14}.
The subgraphs $\{1,2,5\}$ and $\{1,2,3,4\}$ are 1VI.
The subgraph $\{1,2,3,5\}$ is 1VR
and its 1VI components are $\{1,2,5\}$ and $\{3\}$.
The subgraph $\{1,2,3\}$ is 1VR
and its 1VI components are $\{1\}$, $\{2\}$ and $\{3\}$.

A set $f$ of 1VI subgraphs 
is called an {\em ultraviolet (UV) forest} if the
following conditions hold:

({\em i}) for any pair $\gm,\gm'\in f$,
we have either $\gm\subset\gm'$, $\gm'\subset\gm$ or
${\cal L}(\gm\cap\gm')=\emptyset$;
\\\indent
({\em ii}) if $\gm^1,\ldots,\gm^n\in f$
and ${\cal L}(\gm^i\cap\gm^j)=\emptyset$ for any pair from this
family, the subgraph $\cup_i \gm^i$ is 1VR.

In other words, the number of loops in $\cup_i \gm^i$ (where
$\gm^i$ are disjoint with respect to lines and belong to a UV
forest) is equal to the sum of the numbers of loops of  $\gm^i$.
The term ``UV'' is used because the UV divergences are due to the
integration over small values of $\al_l$ where the exponent in
(\ref{alpha}) is irrelevant and they are generated by the
singularities of the factor ${\cal U}^{ -d/2}$.
We are going to show that the resolution of the UV singularities can be performed
by the use of sectors associated with 1VI subgraphs.

For example, the set
$\{1\}$, $\{2\}$, $\{3\}$
of subgraphs of Fig.~\ref{fi14} is a UV forest and $\{1,2,5\}$,
$\{3\}$ is also a UV forest but the set $\{1\}$, $\{2\}$, $\{3\}$,
$\{4\}$ is not a UV forest because the condition ({\em ii})
breaks down.

Let ${\cal F}$ be a {\em maximal} UV forest (i.e. there are no UV
forests that include  ${\cal F}$) of a given graph $\Gm$.
 An element $\gm\in {\cal F}$ is called {\em trivial} if
it consists of a single line and is not a loop line.
Any maximal UV forest
has $h$ non-trivial and $L-h$ trivial elements.

Let us define the mapping
$\sg: \; {\cal F}\rightarrow {\cal L}$ such that $\sg(\gm) \in {\cal L}(\gm)$
and $\sg(\gm) \insl  {\cal L}(\gm')$ for any $\gm'\subset \gm$, $\gm'\in {\cal F}$.
The inverse mapping
$\sg^{-1}: \; {\cal L}\rightarrow{\cal F} $ exists and can be defined as follows:
$\sg^{-1}(l)$ is
the minimal element of the UV forest  ${\cal F}$ that contains the line $l$.
Let us denote by $\gm_+$ the minimal element of  ${\cal F}$ that strictly includes
the given element $\gm$.

 For a given maximal UV forest ${\cal F}$,
let us define the corresponding sector ($f$-sector)  as
\be
{\cal D}_{\cal F} = \left\{ \ual | \al_l\leq \al_{\sg(\gm)},
\; l\in \gm\in {\cal F}\right\}\;.
\ee
The intersection of two
different $f$-sectors is of measure zero; the union of all the
sectors gives the whole integration domain of the alpha
parameters.
 For a given $f$-sector, let us introduce new variables labelled by
the elements of ${\cal F}$,
\be
\al_l  = \prod_{\gm\in {\cal F}:\; l\in\gm} t_{\gm}\;,
\label{S-s-v}
\ee
where the corresponding Jacobian is $\prod_{\gm} t_{\gm}^{L(\gm)-1}$.
The inverse formula is
\be
t_{\gm}= \left\{ \begin{array}{ll}
 \al_{\sg(\gm)}/\al_{\sg(\gm_+)}
& \;\;\mbox{if  $\gm$ is not maximal}  \\
 \al_{\sg(\gm)}  & \;\;\mbox{if $\gm$ is maximal }
\end{array} \right.\;.
\label{S-s-v-back}
\ee

Consider, for example,
the following  maximal UV forest ${\cal F}$ of Fig.~\ref{fi14} consisting of
$\gm^1=\{1\},\; \gm^2=\{2\},\; \gm^3=\{3\},\;
\gm^4=\{1,2,5\},\; \gm^5=\Gm$. The mapping $\sg$ is
$\sg(\gm^1)=1,\; \sg(\gm^2)=2,\;\sg(\gm^3)=3,\;
\sg(\gm^4)=5,\;\sg(\gm^5)=4$.
The sector associated with this maximal UV forest is given by
${\cal D}_{\cal F}=\{ \al_{1,2}\leq \al_5\leq\al_4,\; \al_3\leq\al_4\}$ and
the sector variables are $t_{\gm^1}=\al_1/\al_5,\; t_{\gm^2}=\al_2/\al_5,\;
t_{\gm^3}=\al_3/\al_4,\; t_{\gm^4}=\al_5/\al_4,\; t_{\gm^5}=\al_4$.

All the maximal UV forests of the given graph can be constructed at least in
two ways.

{\em Way~1.}
We imply that the lines are enumerated.
Let us consider the sequence of subgraphs $\gm_l$ consisting of
lines $\{1,2,\ldots,l\}$, respectively, with $l=1,\ldots,L$.
For each $l$, let us take the 1VI component of $\gm_l$
that includes the line $l$. The set of all these components is
a maximal UV forest. Then we construct in a similar way the UV forests
for other $L!-1$ enumerations of the set of lines. After this
we leave only distinct maximal UV forests.

{\em Way~2.}
Since we consider a 1VI graph we include it into any maximal
forest. Let us delete a line from it. The resulting graph is decomposed
as the union of its 1VI components which we include into the
maximal UV forest. Then we continue this process by deleting a
line from some 1VI component which is not a single line, etc.

In the sector corresponding to a given maximal UV forest $f$,
the function ${\cal U}$ takes the form
\be
{\cal U} =  \prod_{\gm\in f}  t_{\gm}^{h(\gm)}
\left[1+P_f\right]
\label{U-fac}
\;,
\ee
where $P_f$ is a non-negative polynomial and the product is over
elements of the given maximal UV forest $f$. We will call such a
factorization {\em proper}.

This factorization formula is proved by constructing an appropriate
tree. One uses the relation
\be
\prod_{l \insl T } \al_l
 =  \prod_{\gm\in f}
t_{\gm}^{h(\gm) + c(\gm\cap T)-c(\gm)} \;,
\label{prodfac}
\ee
where $T$ is a tree or a 2-tree
so that the factorization reduces to
constructing a tree that provides the minimal value
of the non-negative quantity $c(\gm\cap T)-c(\gm)$.
Let $T_0$ be the tree composed of all trivial elements
of the given maximal UV-forest $F$. In other words, this tree can
be constructed as follows. One uses an order of lines which was
used within Way~1 for the construction of the given maximal UV forest
$f$  and includes the given line in the tree if a loop is not generated.
One can obseve that this tree $T_0$ provides the zero value of
$c(\gm\cap T_0)-c(\gm)$ for all the elements of the given maximal forest.

To analyze convergence of the integral (\ref{alpha})  large values of
$\al_l$ (in particular, to reveal infrared (IR)
divergences) one has to take into account the exponent as well. A
possible way is to separate the integration over every $\al_l$
into $(0,1)$ and $(1,\infty)$
and then to deal with each of these $2^L$ regions separately
--- see, e.g. \cite{books1,books1a}. This can be enough for a general analysis but
cannot be good from the practical point of view because the number
of the resulting sectors will be too large.
A more reasonable approach is to turn \cite{Speer2} to an integral with a
compact integration domain, where both UV and IR divergences are
somehow mixed up and manifest themselves as divergences at
small values of parameters of integration. We will do this in the
next section.

\section{Strategies to reproduce Hepp and Speer sectors}

In \cite{BH}, the starting point is the alpha representation
(\ref{alpha-d-mod}) where {\em primary sectors} are introduced.
The set of primary sectors corresponds to the different choices of a line in
the given graph. At this step, one chooses a line $l=1,...,L$ and
defines a sector $\Delta_l$ by $\al_i\leq \al_l,\; i\neq l$ and the sector
variables by $\al_i =  t_i\al_l, \; i\neq l$.
The integration over $\al_l$ is then taken due to the delta function whose
argument is supposed to be the sum of all the variables minus one.

One can also start directly from (\ref{alpha}) and introduce primary sectors
$\al_i\leq \al_l,\; i\neq l$ there. For example, in the case of $l=L$, using
the homogeneity properties of the functions in the representation and
explicitly integrating over $\alpha_L$
we obtain the contribution of $\Delta_L$ as
\bea
F^{(L)}
& =&
 \frac{\left(\I\pi^{d/2} \right)^h
\Gm(a-h d/2)}{\prod_l\Gm(a_l)}
\int_0^1 \ldots\int_0^1  \prod_l^{L-1} \al_l^{a_l-1}\,
\nn \\ && 
\times
\frac{
\hat{\cal U}^{a-(h+1) d/2}}
{\left[-\hat{\cal V} +\hat{\cal U}
\left(\sum_{l=1}^{L-1} m^2_l \prod_{l=l'}^{L-1} \al_{l'}+m^2_L\right)\right]^{a-h d/2}}
\dd \al_1 \ldots \dd \al_{L-1} \;,
\label{primsec}
\eea
where
\be
\hat{\cal U}={\cal U}(\al_1,\ldots,\al_{L-1},1)\;, \;\;\;
\hat{\cal V}={\cal V}(\al_1,\ldots,\al_{L-1},1)\;.
\ee
Without loss of generality let us consider only this primary
sector.

Let us remind that, for the case of non-zero masses, a general
analysis of the factorization is not known even for Euclidean
external momenta. Let us therefore turn to the pure massless case,
as in \cite{Speer2}.
Let us describe how sectors of Speer type can be introduced in
such a way that the whole integrand of (\ref{primsec})
has a proper factorization.
As we could see in the previous section, the use of $f$-sectors provides
a proper factorization (\ref{U-fac}) of the function $\cal U$ so
that the factor $\hat{\cal U}^{a-(h+1) d/2}$ in (\ref{primsec}) is
properly factorized. However, these sectors generally do not
provide a factorization of the second non-trivial factor. This can
be seen using our example of Fig.~\ref{fi14}. We are going to use smaller
sectors which are in fact obtained from the $f$-sectors
generated by the graph $\Gm$ by a further decomposition.

Let  $\Gm^\infty$ be the graph obtained from $\Gm$ by adding a new
vertex $v^\infty$ and connecting it with all the external $n+1$
vertices by additional lines. These lines are only auxiliary and
no propagators
correspond to them. When writing down the function $\cal U$
for $\Gm^\infty$, let us include, by definition, these additional
lines into any tree. Then in the case of two external vertices (i.e. for $n=1$)
we have
\[
{\cal V}_{\Gm}={\cal U}_{\Gm^\infty} q^2
\]
where $q$ is the only external momentum.

Let us define sectors in a way similar to the previous section but using,
instead of 1VI subgraphs, another set of subgraphs which we
call \textit{$s$-irreducible}. If a subgraph $\gm$ does not have all the
external vertices in the same connectivity component and if it is
1VI let us call it $s$-irreducible as well. If a subgraph $\gm$ has all
the external vertices in the same connectivity component let us call
it $s$-irreducible if the graph $\gm^\infty$ is 1VI.
Then the maximal forests consisting of $s$-irreducible subgraphs
can be constructed again by Way~1 or Way~2.

We define sectors (we name them \textit{Speer sectors}) in a way similar to the sectors discussed in the
previous section.
We introduce sector variables by the same formula (\ref{S-s-v})
as above. The factorization of the function $\cal V$ follows
from its definition (\ref{Aform}) and the auxiliary relation
(\ref{prodfac}). The 2-tree that provides a minimal value of
the non-negative quantity $c(\gm\cap T)-c(\gm)$
can be constructed by a procedure similar to the procedure used
for the function $U$: one considers the  lines in the order used
for the construction of the given $f$-forest by Way~1
and includes the given line
into the 2-tree if a loop is not generated {\em and} if this is
not the line whose inclusion would connect all the external vertices.

By construction, for such a 2-tree $T_0$, we obtain
$c(\gm\cap T_0)-c(\gm)=\theta(\gm)$ where $ \theta(\gm)=1$
if the external vertices are connected in
$\gm$ and $\theta(\gm)=0$ otherwise. Hence we obtain a
proper factorization
\be
{\cal V} =  \prod_{\gm\in f}  t_{\gm}^{h(\gm)+\theta(\gm)}
\left[q^2_{T_0}+P_{\cal V}\right]
\label{V-fac}
\;,
\ee
where $P_{\cal V}$ is a non-negative polynomial.

Obviously, the Speer sectors can be obtained from those associated
with the graph $\Gm$ by a further decomposition, so that the
factorization of the function ${\cal U}$ in the corresponding
variables also holds and has the form similar to (\ref{U-fac})
with the same exponents.

Let us turn to the modern strategies of sector decompositions.
After introducing primary sectors, one obtains the contribution
(\ref{primsec}) and other $L-1$ contributions of the same type.
At each step, one chooses a subset of the indices
$\nu=\{i_1,\ldots,i_k\}$ and an index $i_r$ from this subset and
defines a sector $\al_i\leq \al_{i_r},\; i\neq i_r$ and the
sector variables by $\al_i = t_i \al_{i_r} , \; i\neq i_r$
To formulate a strategy of introducing iterative sectors one needs to fix
rules for determining subsets $\nu$ at every recursive step
\footnote{Some strategies use more general sectors by
comparing the integrations variables in different powers.}.

The first know sector decomposition strategy is described in \cite{BH};
three strategies guaranteed to terminate ($A$, $B$ and $C$) and one 
strategy not guaranteed to terminate ($X$) are presented in \cite{BognerWeinzierl};
they all are based on analyzing the functions $\cal U$ and
$\cal V$ and choosing a subset of indices depending on their properties. Strategy $S$
\cite{FIESTA} is a bit different and is based on analyzing the polytopes of
weights and their lowest faces.

With the Hepp sectors, the situation is obvious: they are reproduced when we
consider {\em maximal} subsets of lines at each step, i.e. with one line less
than before this.

To reproduce the choice of Speer sectors within a sector
decomposition let us remind the Way~2 to construct sectors.
One has to consider only subsets of the indices
$\nu=\{i_1,\ldots,i_k\}$ that correspond to $s$-irreducible subgraphs.

\begin{figure}[h]
\begin{center}
\includegraphics[width=.5\textwidth]{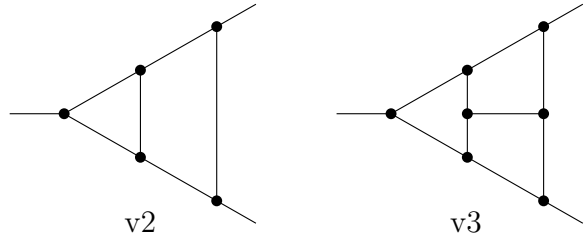}
\caption[]{Vertex off-shell diagrams}
\label{v2-v3}
\end{center}
\end{figure}

\begin{figure}[hbt]
\begin{center}
\includegraphics[width=.8\textwidth]{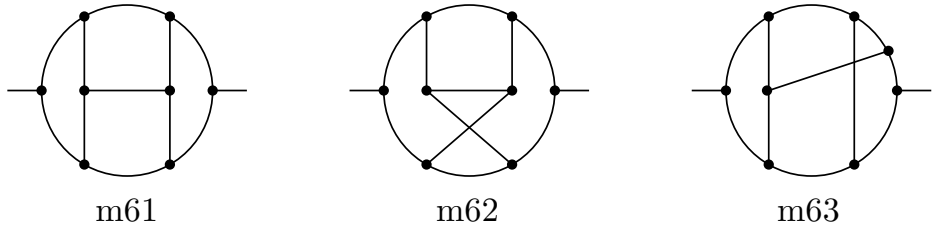}
\caption[]{Most complicated four-loop propagator diagrams}
\label{m61-63}
\end{center}
\end{figure}

We compared the number of sectors for numerous examples and discovered
a curious phenomenon: the set of the sectors within Strategy~S and
the set of the Speer sectors is always the same.
In particular, this is observed in the two examples of massless vertex diagrams
shown in  Fig.~\ref{v2-v3} at Euclidean external momenta,
and in rather non-trivial examples of three four-loop
propagators diagrams of Fig.~\ref{m61-63}
which are the most complicated master
integrals among all four-loop massless propagators integrals\footnote{
Analytical results in expansion in $\ep$ up to $\ep^0$ for these diagrams will be
published soon \cite{BaCh}.}.
\begin{figure}[hbt]
\begin{tabular}{|c|l|l|l|l|}
  \hline
  diagram & S  & X \\
  \hline
 v2 &  102    &  102  \\
  v3 &  2160  & 2251 \\
  m61 &  26208   & 32620 \\
  m62 & 26304   & 27540 \\
  m63 &  27336   & failed \\
  \hline
\end{tabular}
\end{figure}
The results of this analysis are shown in Table~1 where the number of the sectors
is shown. The first column stands for Strategy~S and Speer sectors; the resulting
number of the sectors is the same.
For comparison, we included the second column for Strategy~X
\cite{BognerWeinzierl} which has been proved to be very effective in a number
of complicated calculations. Here `failed' means that a factorization by sector
decomposition has not been achieved for a reasonable amount of time (at least
not for one day.)

\section{Conclusion}

In fact, our motivation to recall Speer sectors was to
suggest to use them within {\tt FIESTA} since they are optimal
for Feynman integrals at all Euclidean
external momenta. However it turned out that these sectors are
reproduced, in rather nontrivial examples, within Strategy~S.
Therefore we can conclude that Strategy~S has chances to be an
optimal universal strategy which always terminates and provides a
proper factorization.

Let us remind that Speer sectors have a certain physical meaning:
the integration over the sector variable $t_{\gm}$ is responsible for
UV divergences and, if $\gm$ contains all the external vertices,
for off-shell IR divergences. If on-shell and collinear divergences are present,
Speer sectors are no longer applicable. Then one can use sector decompositions,
e.g. within Strategy~S, and this strategy could help to reveal the physical
meaning of the sectors. To do this one might start with analyzing simple typical
diagrams with on-shell or/and collinear divergences.

\vspace{0.2 cm}

{\em Acknowledgments.}
We are grateful to G.~Heinrich for helpful discussions. 
Many thanks to N.~Glover for kind hospitality during our visit to the 
Institute of Particle Physics Phenomenology in Durham, where a 
part of this work was done.
The work was supported by the Russian Foundation for Basic
Research through grant 08-02-01451.

\end{document}